\documentstyle[12pt]{article}

\title{
\begin{flushright}
{\normalsize Yaroslavl State University\\
             Preprint YARU-HE-96/05\\
             hep-ph/9612312} \\[5mm]
\end{flushright}
The neutrino energy and momentum loss 
through the process $\nu \to \nu e^- e^+$
in a strong magnetic field}

\author{A.V.~Kuznetsov\thanks{E-mail address:avkuzn@univ.uniyar.ac.ru}, 
N.V.~Mikheev\thanks{E-mail address:
mikheev@yars.free.net}\\
{\small\it Division of Theoretical Physics, Department of Physics,}\\
{\small\it Yaroslavl State University, Sovietskaya 14,}\\
{\small\it 150000 Yaroslavl, Russian Federation.}}

\date{10 December 1996}

\begin{document}

\maketitle

\begin{abstract}
A process of the electron--positron pair production by neutrino 
propagating in a strong magnetic field is investigated in the framework 
of the Standard Model. The process probability and the mean values 
of the neutrino energy and momentum loss are calculated. 
Possible astrophysical manifestations of the process considered 
are briefly analysed.
\end{abstract}

\vglue 5mm

\begin{center}
{\it Submitted to Physics Letters B} 
\end{center}

\newpage

The neutrino interaction with an intensive electromagnetic field plays 
an important role in neutrino physics. The interaction was generally 
considered beyond the Standard Model, for example, due to an unusually 
large neutrino magnetic moment~\cite{Okun}. In the recent times an 
interesting effect of significant enhancement of the probability of the 
neutrino radiative decay $\nu_i \to \nu_j \gamma$ by a magnetic field 
(magnetic catalysis) was discovered in the frame of the Standard Model 
with lepton mixing~\cite{GMV}. 

Another decay channel also exists, $\nu_i \to \nu_j e^- e^+$ 
($i$ and $j$ enumerate the neutrino mass eigenstates but not the neutrinos 
with definite flavors), which is
forbidden in vacuum when $m_i < m_j + 2m_e$. 
However, the kinematics of a charged particle 
in a magnetic field is that which allows to have 
a sufficiently large space--like total momentum for 
the electron--positron pair, and this process is possible even for very light 
neutrinos. It means that a flavor of the ultrarelativistic neutrino is 
conserved in this transition in a magnetic field, to the terms of the order 
of $m^2_{\nu}/E^2_{\nu}$ regardless of the lepton mixing angles. 
Consequently, a question of neutrino 
mixing is not pertinent in this case and the process 
$\nu \to \nu e^- e^+$ can be considered in the frame of 
the Standard Model without lepton mixing.

In this paper we investigate the process $\nu \to \nu e^- e^+$ for the high 
energy neutrino, $E_{\nu} \gg m_e$, in a strong constant magnetic field. 
We consider a magnetic field as the strong one 
if it is much greater than the known Schwinger value 
$B_e = m^2_e/e \simeq 4.41 \cdot 10^{13} G$. 
This process could be of importance in astrophysical applications, e.g. in an 
analysis of cataclysms like a supernova explosion or a coalescence of 
neutron stars, where the strong magnetic fields can exist, and where 
neutrino processes play the central physical role. 
In the present view, the magnetic field strength inside the astrophysical 
objects in principle could be as high as 
$10^{16} - 10^{18} G$, both for toroidal~\cite{Rud,Lip} and for
poloidal~\cite{Bocq} fields. The magnetic fields of the order of
$10^{12} - 10^{13} \, G$ have been observed at the surface of pulsars are 
the so-called `old' magnetic fields and they do not provide a determination 
of the field strength at a moment of the cataclysm~\cite{Lip}.
Curiously, a model was proposed recently for soft gamma-ray repeaters 
to be associated with young neutron stars, with surface dipole fields 
$B \sim 10^{14} - 10^{15} G$ \cite{Thomp}.

Here we calculate the probability of the process and the mean 
values of the neutrino energy and momentum loss through the production of 
electron--positron pairs. 

If the momentum transferred is relatively small, $|q^2| \ll m^2_W$
\footnote{As the analysis shows, it corresponds in this case to the 
neutrino energy $E \ll m^3_W/e B$.}, the 
weak interaction of neutrinos with electrons could be described in the local 
limit by the effective Lagrangian of the form

\begin{equation}
{\cal L} \, = \, \frac{G_F}{\sqrt 2} 
\big [ \bar e \gamma_{\alpha} (g_V - g_A \gamma_5) e \big ] \,
\big [ \bar \nu \gamma^{\alpha} (1 - \gamma_5) \nu \big ] \,,
\label{eq:L}
\end{equation}

\noindent where

\begin{equation}
g_V = \pm {1 \over 2} + 2 sin^2 \theta_W \, , \quad g_A = \pm {1 \over 2}.
\label{eq:gvga}
\end{equation}

\noindent Here the upper signs correspond to the electron neutrino 
($\nu = \nu_e$) when both neutral and charged current interaction takes part 
in a process. The lower signs correspond to $\mu$ and $\tau$ neutrinos 
($\nu = \nu_{\mu}, \nu_{\tau}$), when the neutral current interaction 
is only presented in the Lagrangian~(\ref{eq:L}). 

An amplitude of the process $\nu \to \nu e^- e^+$ could be immediately 
obtained from the Lagrangian~(\ref{eq:L}) where the known solutions of the 
Dirac equation in a magnetic field should be used. 
The general expression for the probability takes a rather complicated form. 
We present here the results of our calculations in two limiting cases which 
have a clear physical meaning: $e B > E^2$ and $e B \ll E^2$. 

In the case when the field strength $B$ appears to be 
the largest physical parameter, the electron and the positron could be born 
only in the states corresponding to the lowest Landau level. Being integrated 
over the $e^- e^+$ pair phase space with the fixed total pair momentum $q$, 
the process probability could be written in the form 

\begin{equation}
w E \, = \, \frac{G_F^2 m^2 e B}{32 \pi^4} \, \int \frac{d^3 k}{\omega} 
\, \frac{e^{-{q^2_{\perp}}/{2 e B}}}
{(q^2_{\parallel})^{3/2} (q^2_{\parallel} - 4 m^2)^{1/2}}\,
\left | g_V (q \tilde \varphi j) -
 g_A (q \varphi \varphi j) \right |^2 \, .
\label{eq:wE1}
\end{equation}

\noindent Hereafter $p^{\alpha} = (E,\vec p), \; k^{\alpha} = 
(\omega, \vec k)$ are the four-momenta of the initial and the final neutrinos, 
respectively; $m$ is the electron mass; $q = p - k$;  
$j^{\alpha} = \bar \nu(k) \gamma^{\alpha} 
(1 - \gamma_5) \nu(p)$; $(q \varphi \varphi j) = q^{\alpha} \varphi_{\alpha 
\beta} \varphi^{\beta \delta} j_{\delta}$, 
$\varphi_{\alpha \beta}$ is the dimensionless external field tensor,
$\varphi_{\alpha \beta} = F_{\alpha \beta}/B$,
$\tilde \varphi_{\alpha \beta} = {1 \over 2} \varepsilon_{\alpha \beta \rho 
\sigma} \varphi^{\rho \sigma}$ is the dual tensor; 
$q^2_{\parallel} = (q \tilde \varphi \tilde \varphi q)$, 
$q^2_{\perp} = (q \varphi \varphi q)$. 

It should be noted that an expression for the probability of the process 
and in part Eq.~(\ref{eq:wE1}) could be obtained by another easier 
way. As is well known, the cross-section of the reaction 
$\nu \tilde \nu \to e^- e^+$ is connected with the imaginary part of the 
amplitude of the $\nu \tilde \nu \to \nu \tilde \nu$ transition via the 
electron loop (see Fig.1) by the unitarity condition  

\begin{equation}
\sigma(\nu \tilde \nu \to e^- e^+) = {1 \over q^2} \,
Im \,{\cal M}(\nu \tilde \nu \to \nu \tilde \nu) .
\label{eq:unit}
\end{equation}

\noindent One can easily see that Eq.~(\ref{eq:unit}) allows to find 
a probability of the crossed channel of the reaction, namely, the process 
 $\nu \to \nu e^- e^+$, by integrating over the phase space of the final 
neutrino

\begin{equation}
w(\nu \to \nu e^- e^+) E \, = \, \frac{1}{16 \pi^3} \, \int 
\frac{d^3 k}{\omega} 
\, Im \,{\cal M}(\nu \tilde \nu \to \nu \tilde \nu) .
\label{eq:wE2}
\end{equation}

\noindent Within the above--mentioned limiting case $e B > E^2$ 
an imaginary part of the amplitude, see Fig.1, corresponds to the 
virtual electron and positron to become on-shell in the lowest Landau level. 
A straightforward calculation of Eq.~(\ref{eq:wE2}) leads to 
Eq.~(\ref{eq:wE1}) immediately. 

Integrating Eq.~(\ref{eq:wE1}) over the phase space, one obtains the following 
expression for the probability  

\begin{eqnarray}
w(\nu \to \nu e^- e^+) & =& \frac{G_F^2 (g_V^2 + g_A^2)}{16 \pi^3} \, 
e B E^3 sin^4 \theta \cdot f_1 (\lambda),
\label{eq:wE3}\\
f_1 (\lambda)\, &=& 1 \,-\, {2 \over 3} \lambda \,+ \, {5 \over 16} \lambda^2 
\, - \, {7 \over 60} \lambda^3 \, + \dots ,
\end{eqnarray}

\noindent 
where $\lambda = (p \varphi \varphi p)/e B =E^2 sin^2 \theta/e B$, 
$\theta$ is an angle between the initial neutrino momentum
$\vec p$ and the magnetic field direction. 

In another limiting case, $E^2 \gg e B$, when a great number of the 
Landau levels could be excited, our result can also be represented in a 
simple form. Using the Eq.~(\ref{eq:wE2}) we obtain the following expression 
for the process probability

\begin{equation}
w(\nu \to \nu e^- e^+) E \, = \, \frac{G_F^2 (g_V^2 + g_A^2)}{27 \pi^3} \, 
m^6 \chi^2 \, (ln \chi - {1 \over 2} ln 3 - \gamma_E - {29 \over 24}) ,
\label{eq:wE4}
\end{equation}

\noindent where $\chi = e (pFFp)^{1/2}/m^3 = \beta E sin \theta/m$ 
is the so-called field dynamical parameter, $\beta = B/B_e \gg 1$, 
$\gamma_E = 0.577 \dots$ is the Euler constant. 

A probability of this process in the last limiting case was obtained earlier 
in the paper~\cite{Bor} by calculating of an imaginary part of the two-loop 
amplitude of the $\nu_{\mu} \to \nu_{\mu}$ transition via the virtual 
$e^- e^+$ pair. However, some 
numerical errors were committed there. In particular, 
an extra factor of $1 \over 2$ was inserted. Besides, the calculation 
technique used in Ref.~\cite{Bor} allows to find an integral probability 
of the process $\nu \to \nu e^- e^+$ only, but the final neutrino spectrum 
could not be obtained. Thus, it would not be possible to investigate some 
interesting astrophysical consequences of this process to be discussed below. 

It should be noted that a practical significance of this process for 
astrophysics could be in the mean values of the neutrino energy and momentum 
loss rather than in the process probability~(\ref{eq:wE3}), (\ref{eq:wE4}). 
These mean values could be found from the four-vector 

\begin{equation}
Q^{\alpha} \, = \, \frac{1}{16 \pi^3} \, \int 
\frac{d^3 k}{\omega} \, q^{\alpha} 
\, Im \,{\cal M}(\nu \tilde \nu \to \nu \tilde \nu) .
\label{eq:Q}
\end{equation}

\noindent Its zero component is connected with the mean neutrino energy 
loss in a unit time, ${\cal I} = Q_0/E$. The space components of the 
four-vector~(\ref{eq:Q}) are connected similarly with the neutrino momentum 
loss in unit time, $\vec {\cal F} = \vec Q/E$. Here we present the expressions 
for $Q^{\alpha}$ in two limiting cases used above for the 
probability:

i) $e B \sim E^2$

\begin{eqnarray}
Q^{\alpha} & = & \frac{G_F^2 e B (p \varphi \varphi p)^2}{48 \pi^3} 
\{(g_V^2 + g_A^2)[p^{\alpha} \cdot f_2 (\lambda) - 
2 (\varphi \varphi p)^{\alpha} \cdot f_3 (\lambda)] 
\nonumber \\
& + & 2 g_V g_A (\tilde \varphi p)^{\alpha} \cdot f_2 (\lambda) \} ,
\label{eq:Q1}\\
f_2 (\lambda)\, &=& 1 \,-\, {5 \over 8} \lambda \,+ \, {21 \over 80} \lambda^2 
\, - \, {7 \over 80} \lambda^3 \, + \dots ,
\nonumber \\
f_3 (\lambda)\, &=& 1 \,-\, {15 \over 16} \lambda \,+ \, {21 \over 40} \lambda^2 
\, - \, {7 \over 32} \lambda^3 \, + \dots ;
\nonumber 
\end{eqnarray}

ii) $E^2 \gg e B$

\begin{eqnarray}
Q^{\alpha} \,& =& \, {7 \over 16} \, \frac{w E}{ln \chi - 2.335} \,
[\,p^{\alpha} (ln \chi - 1.888) - \sqrt{3} \, \frac{\beta^2}{\chi} \,
(\varphi \varphi p)^{\alpha} 
\nonumber \\
&-& 7.465 \, \frac{g_V g_A}{g_V^2 + g_A^2} \, \frac{\beta}{\chi^{2/3}} \,
(\tilde \varphi p)^{\alpha}],
\label{eq:Q2}
\end{eqnarray}

\noindent 
where the probability $w$ should be taken from Eq.~(\ref{eq:wE4}).
Eqs.~(\ref{eq:wE3}), (\ref{eq:wE4}), (\ref{eq:Q1}), 
(\ref{eq:Q2}) are also applicable for the process with antineutrino 
$\tilde\nu \to \tilde\nu e^- e^+$ due to the $CP$--invariance of the weak 
interaction.

Let us note that our results are valid in the presence of plasma with the 
electron density $n \sim 10^{33} - 10^{34} cm^{-3}, 
\, (n = n_{e^-} - n_{e^+})$. This is due to a peculiarity of the statistics 
of the relativistic electron gas in a magnetic field~\cite{Lan}.
The chemical potential in a very strong field
 
\begin{equation}
\mu \, = \, \frac{2 \pi^2 n}{e B} \,\simeq \, 
 0.26 MeV 
\left (\frac{n}{10^{33} cm^{-3}} \right )
\left (\frac{10^{17}G}{B} \right ) 
\label{eq:mu}
\end{equation}

\noindent is significantly less than in the case without field. 
Because of this, 
the suppressing statistical factors do not arise in integrating over the 
phase space of the $e^- e^+$ pair.

To illustrate the formulae obtained we shall consider 
the astrophysical process of a birth of the magnetized neutron star, 
pulsar, for example, in a supernova explosion. 
Let us suppose that in the cataclysm a very strong magnetic field of the order 
of $10^{16} - 10^{18} G$~\cite{Rud, Bocq, Lip} arises in any reason in the 
vicinity of a neutrinosphere. The electron density in this region will 
be considered to be not too high, so the electron gas could be treated as 
the nondegenerated one. In this case the neutrino propagating through the 
magnetic field will loose the energy and the momentum in accordance with 
our formulas. A part of the total energy lost by neutrinos in the field 
as well as the total momentum received by a rest of the star from outgoing 
neutrinos could be estimated from Eq.~(\ref{eq:Q1}):

\begin{equation}
\frac{\Delta {\cal E}}{{\cal E}_{tot}} 
 \, \sim \, 2 \cdot 10^{-2}\, 
\left (\frac{B}{10^{17}G} \right )\,
\left (\frac{\bar E}{10 MeV}\right )^3\,
\left (\frac{ \Delta \ell}{10 km}\right ), 
\label{eq:E} 
\end{equation}

\begin{eqnarray}
\vec {\cal P}& =& (0, 0, {\cal P}), 
\nonumber \\
{\cal P} & \sim & 3 \cdot10^{40} \frac{g\,cm}{s} 
\left (\frac{{\cal E}_{tot}}{4 \cdot 10^{53} erg} \right )
\left (\frac{B}{10^{17}G} \right )
\left (\frac{\bar E}{10 MeV}\right )^3 
\left (\frac{ \Delta \ell}{10 km}\right ),
\label{eq:P}
\end{eqnarray}

\noindent here the $z$ axis is directed along the magnetic field,
$\Delta \ell$ is a characteristic size of the region where the field 
strength varies insignificantly.
${\cal E}_{tot}$ is the total energy carried off by
neutrinos in a supernova explosion, 
$\bar E$ is the neutrino energy averaged over the neutrino spectrum.
Here we take the energy scales which are believed to be typical for 
supernova explosions~\cite{Imsh}.
One can see from Eq.~(\ref{eq:E}) that the effect could 
manifest itself at a level of about percent. In principle, it could be 
essential in a detailed theoretical description of the process of 
supernova explosion. 

Let us note that an origin of the asymmetry of the neutrino momentum loss 
with respect to the magnetic field direction is a manifestation 
of the parity violation in weak interaction, because the ${\cal P}$ value 
is defined by the term in Eq.~(\ref{eq:Q1}) proportional to the product of 
the constants $g_V$ and $g_A$. This asymmetry results in the recoil  
``kick'' velocity of a rest of the cataclysm. 
If the physical parameters would have in any reason the values of order of 
the scales in Eq.~(\ref{eq:P}), it would provide a ``kick'' velocity 
of order 150 km/s for a pulsar with mass of order of the solar mass. 

\medskip

We are grateful to V.M.~Lipunov for helpful discussions, and 
to A.A.~Gvozdev and L.A.~Vassilevskaya for useful remarks.

The work of N.V.~Mikheev was supported in part by a Grant N d104
from the International Soros Science Education Program.


\unitlength=0.75mm
\special{em:linewidth 0.4pt}
\linethickness{0.4pt}
\hspace{-10mm}
\begin{center}
\begin{picture}(50.00,57.00)
\put(30.00,38.00){\circle{16.00}}
\put(30.00,38.00){\circle{13.00}}
\put(22.50,38.00){\circle*{2.50}}
\put(37.50,38.00){\circle*{2.50}}

\put(37.50,38.00){\line(2,1){20.00}}
\put(37.50,38.00){\vector(2,1){15.00}}
\put(37.50,38.00){\line(2,-1){20.00}}
\put(37.50,38.00){\vector(2,-1){15.00}}

\put(22.50,38.00){\line(-2,1){20.00}}
\put(8.50,45.00){\vector(2,-1){1.00}}
\put(22.50,38.00){\line(-2,-1){20.00}}
\put(8.50,31.00){\vector(2,1){1.00}}

\put(30.00,26.00){\makebox(0,0)[cc]{$e$}}
\put(10.00,26.00){\makebox(0,0)[cc]{$\nu$}}
\put(50.00,26.00){\makebox(0,0)[cc]{$\nu$}}
\put(30.00,50.00){\makebox(0,0)[cc]{$e$}}
\put(10.00,50.00){\makebox(0,0)[cc]{$\tilde \nu$}}
\put(50.00,50.00){\makebox(0,0)[cc]{$\tilde \nu$}}
\put(30.00,15.00){\makebox(0,0)[cc]{\large Fig.~1.}}
\end{picture}
\end{center}



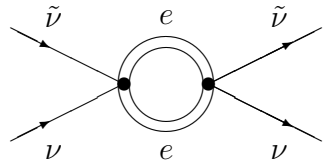
Fig.1. The Feynman diagram for the process
$\nu \tilde \nu \to \nu \tilde \nu$. The double line corresponds 
to the exact propagator of an electron in a magnetic field.


\end{document}